\documentclass[prd,12pt,aps,onecolumn,tightenlines,showpacs,floatfix,amssymb,amsmath]{revtex4-2}
\usepackage{slashed}
\usepackage{siunitx}
\usepackage{epsfig}
\usepackage[colorlinks=true]{hyperref}
\hypersetup{
    bookmarks=true,         % show bookmarks bar?
    colorlinks=true,       % false: boxed links; true: colored links
    linkcolor=black,          % color of internal links (change box color with linkbordercolor)
    citecolor=black,        % color of links to bibliography
    filecolor=black,         % color of file links
    urlcolor=blue        % color of external links
}

\newcommand{\beqn}{\begin{eqnarray}}
\newcommand{\eeqn}{\end{eqnarray}}
\newcommand{\beqs}{\begin{subequations}}
\newcommand{\eeqs}{\end{subequations}\\[-2mm]\noindent}
\newcommand{\eq}[1]{(\ref{#1})}

\newcommand{\bs}{\boldsymbol}

\usepackage{color}
\definecolor{purple}{rgb}{0.8,0,0.6}

\newcommand{\add}[1]{#1}

\begin{document}

\title{Rotational diode: Clockwise/counterclockwise asymmetry \\ in conducting and mechanical properties of rotating (semi)conductors}

\author{M. N. Chernodub}
\affiliation{Institut Denis Poisson UMR 7013, Universit\'e de Tours, 37200 France}
\affiliation{Pacific Quantum Center, Far Eastern Federal University, Sukhanova 8, Vladivostok, 690950, Russia}

\begin{abstract}
We argue that certain materials exhibit asymmetry of their mechanical and conducting properties with respect to clockwise/counterclockwise rotation. We show that a cylinder made of a suitably chosen semiconductor coated in a metallic film and placed in the magnetic-field background can serve as a ``rotational diode'' which conducts electricity only at a specific range of angular frequencies. The critical angular frequency and the direction of rotation can be tuned with the magnetic field's strength. Mechanically, the rotational diode possesses different moments of inertia when rotated in clockwise and counterclockwise directions. These effects emerge as a particularity of the Fermi-Dirac statistics of electrons in rotating conductors. 
\end{abstract}


\date{\today}

\maketitle

\paragraph*{Introduction.} Our daily-life experience tells us that any physical body has the same moment of inertia with respect to rotations in clockwise and counterclockwise directions. In our paper, we show that this statement is no more correct at the quantum level if the statistical quantum effects of electronic systems are taken into account. 

Effects of gravity, rotation, and acceleration on electromagnetic, thermodynamic, and transport properties of physical systems have been a subject of intense interest throughout decades~\cite{ref:Darling}. The Einstein--de Haas~\cite{ref:EdH} and Barnett\cite{ref:Barnett} effects relate mechanical torque and magnetization in ferromagnets. In metals, the uniform rotation acts on electrons via a centrifugal force that produces a small, but experimentally observable radial gradient of electric potential~\cite{ref:Beams}. The proposed rotational analogue of the classical Hall effect~\cite{ref:Ahmedov} highlights a well-known similarly of rotation with the magnetic field in non-relativistic systems. On the quantum level, rapidly rotating C${}_{60}$ fullerenes are suggested to exhibit the Zeeman splitting in energy levels in the absence of a true magnetic field~\cite{ref:C60}.

Accelerating conductors generate intrinsic electric fields~\cite{ref:Moorhead} while gravity exerts a force on the electrons that induces the electric field outside a metal surface~\cite{ref:Witteborn}. Gravitational forces are expected to lead to various thermo-electromagnetic effects in (super)conductors~\cite{ref:Anandan}. At the same time, the quantum Hall conductance, as a true topological quantity, turns out to be insensitive to background gravity~\cite{ref:Obukhov}. The inclusion of the spin degrees of freedom -- and the ability to mechanically manipulate them in noninertial frames -- is expected to play an important role in nano-electromechanical systems within the scope of the rapidly developing field of spintronics~\cite{ref:Matsuo}. The emergence of synthetic gravitational fields in various condensed matter systems opens a new door for discoveries of novel quantum gravito-electromagnetic effects~\cite{ref:Maria,ref:GEV}.

In our paper, we explore the mechanical and transport properties of rotating semiconductors and show that they break the equivalence of clockwise/counterclockwise rotations, which is naively expected for any isolated system. This purely quantum effect has its roots in a simple problem of classical electrodynamics \add{which addresses an interplay between the rotation and the magnetism~\cite{ref:origin}. Before proceeding further, we mention that our discussion} has no direct relation to the Einstein--de Haas effect~\cite{ref:EdH} (which demonstrates the appearance of a mechanical torque exerted by an external magnetic field on a ferromagnet) and the Barnett effect~\cite{ref:Barnett} (which reveals a reciprocal phenomenon: a mechanical rotation changes the magnetization of a spinning ferromagnet). These phenomena appear naturally as a consequence of the conservation of angular momentum. They demonstrate a close relationship between the magnetism, induced by the spin and the orbital motion of the electrons, and the mechanical rotation. Both the Einstein--de Haas and Barnett effects are odd under the time reversal transformation thus \add{maintaining the symmetry of the system under clockwise/counterclockwise flip of the rotation sense} (see, for example, the experimental work~\cite{ref:T:parity}). On the contrary, we consider an effect that breaks the clockwise/counterclockwise symmetry of rotation.

\begin{figure}[!htb]
\begin{center}
\begin{tabular}{cc}
\includegraphics[width=30mm, clip=true]{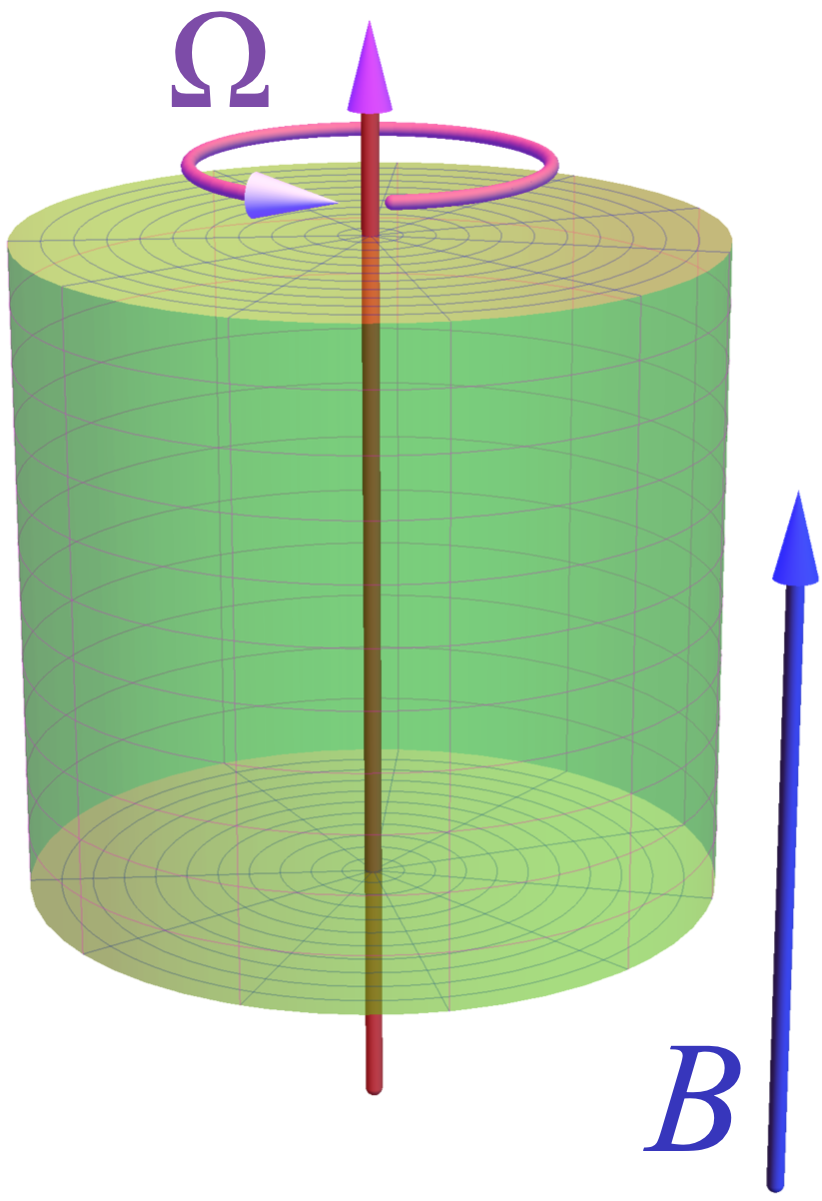}  \\[-46mm]
& \hskip 25mm \includegraphics[width=72mm, clip=true]{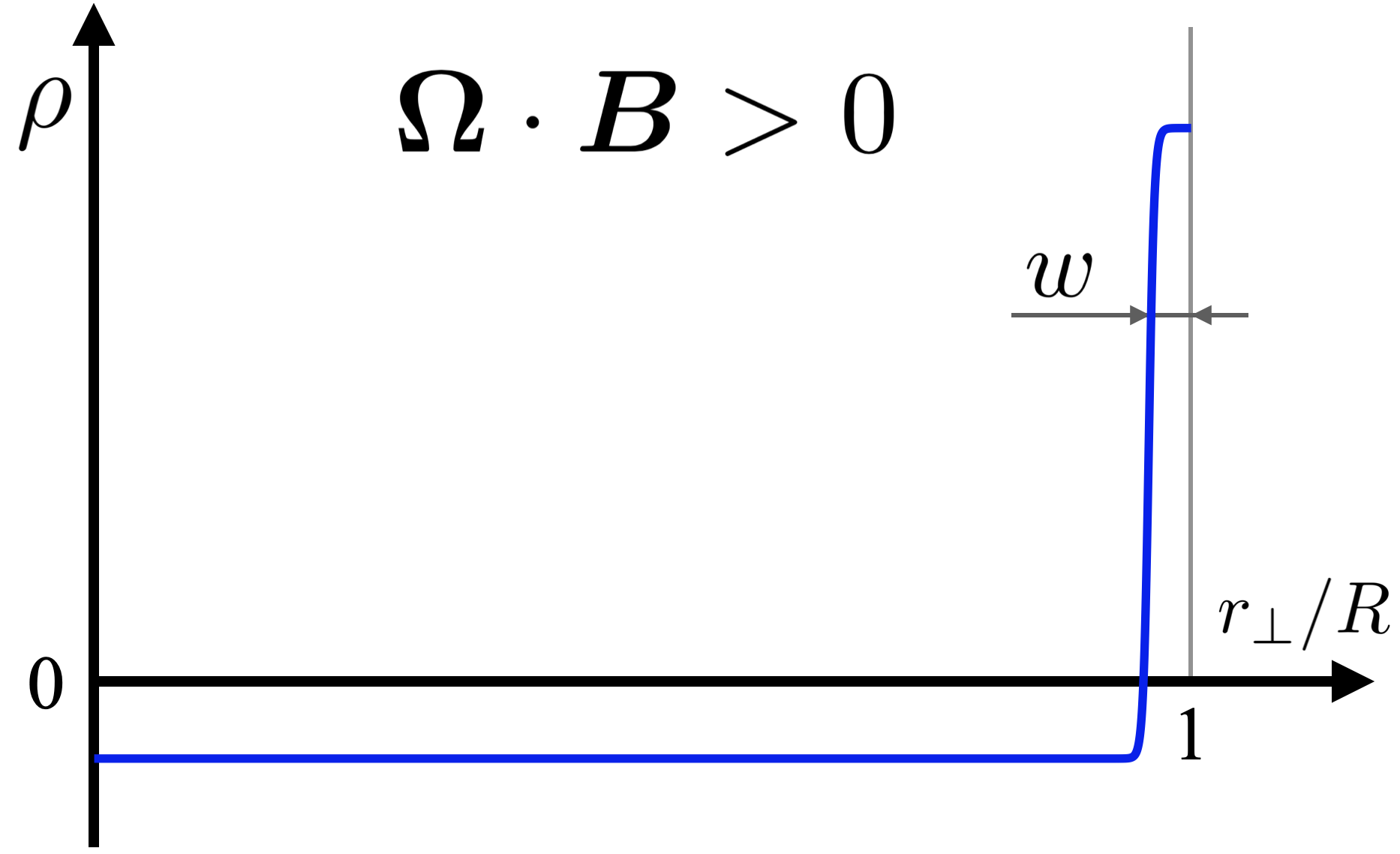} \\[5mm]
(a) & \hskip 5mm (b) 
\end{tabular}
\end{center}
\caption{(a) Rotating conducting (metallic) cylinder in the magnetic field background. (b) Qualitative behavior of the electric charge density inside the cylinder (with $w \ll R$).}
\label{fig:conductor}
\end{figure}

\vskip 1mm
\paragraph*{\add{Rotating conductor in magnetic-field background in classical electrodynamics.}}
Let us consider an uncharged conducting cylinder of radius~$R$ and height~$L$ which rotates rigidly with constant angular velocity ${\bs \Omega} = \Omega \, {\mathrm {\bf e}}_z$ about its symmetry axis~$z$. We place the cylinder in the background of constant and uniform magnetic field 
${\bs B} = B \, {\mathrm {\bf e}}_z$ directed along the axis of rotation as shown in Fig.~\ref{fig:conductor}(a). We set the relative permeability and permittivity of the material to unity $(\epsilon = 1, \, \mu = 1)$ and use Gaussian units.

Treating the system in the scope of classical electrodynamics, one can show that the interior of the cylinder accumulates a uniformly distributed electric charge with bulk density:~\cite{ref:origin}
\beqn
\rho_{\mathrm{bulk}} = - \frac{\Omega B}{2\pi c}\,,
\label{eq:rho:bulk}
\eeqn
while the cylinder boundary (the part tangential to the axis $z$) acquires the uniform surface charge density:
\beqn
\rho_{\mathrm{surf}} = \frac{R \Omega B}{4 \pi c}\,.
\label{eq:rho:boundary}
\eeqn
Since the net electric charge of the isolated cylinder is zero, the bulk~\eq{eq:rho:bulk} and surface~\eq{eq:rho:boundary} charges compensate each other exactly:
$\pi R^2 L \rho_{\mathrm{bulk}}  + 2 \pi R L \rho_{\mathrm{surf}}  = 0$. We do not restrict the mutual directions of ${\bs \Omega}$ and ${\bs B}$ so that the excess of the charge density~\eq{eq:rho:bulk} [and,  respectively, \eq{eq:rho:boundary}] can take both positive and negative values.

The effect originates from a finite conductivity $\sigma \neq 0$ of the rotating cylinder. In an equilibrium state of an isolated physical body, the Joule losses should be absent. This property immediately implies the absence of any dissipative electric currents in the system. In turn, the Ohmic dissipation is generated by a local electric current with respect to the ionic crystal lattice of the conductor. Therefore, the current should vanish in the corotating frame in which the conductor appears static. If the axis of magnetic field and the angular velocity vector are aligned with each other, the local magnetic flux piercing the conductor is not affected by the uniform mechanical rotation. Consequently, the eddy (Foucault) currents and the associated energy losses are absent.

Denoting the quantities in the corotating (laboratory) frame by tilted (non-tilted) variables, the infinitesimal transformation between the coordinates in these frames reads as follows:
\beqn
d {\tilde {\bs r}} = d {\bs r}  - {\bs v}  d t, \qquad d {\tilde t} = d t, \qquad {\bs v} = {{\bs \Omega} \times {\bs r}}\,,
\label{eq:velocity}
\eeqn
where ${\bs v}$ is the local velocity of the fixed point ${\bs r}$ of the cylinder with respect to the laboratory frame. We consider a non-relativistic rotation which guarantees the validity of the causality constraint, $|\Omega| R \ll c$, where $c$ is the speed of light. 

The electromagnetic fields in the laboratory and corotating frames are related to each other as follows:
\beqn
{\tilde {\bs B}} = {\bs B}, 
 \qquad
{\tilde {\bs E}} = \gamma \left({\bs E} + \frac{{\bs v}}{c} \times {\bs B} \right),
 \eeqn
where $\gamma = 1/\sqrt{1 - v^2/c^2}$ is the relativistic Lorentz factor. Hereafter, we ignore all relativistic corrections because they are negligibly small in this system. The rotational motion of the bulk~\eq{eq:rho:bulk} and surface~\eq{eq:rho:boundary} electric charges produces, via the Amp\`ere law, an additional magnetic field which, however, will be dropped out in the following as it gives a tiny correction to the existing magnetic-field background $\bs B$.
Up to negligible relativistic effects, the densities \eq{eq:rho:bulk} and \eq{eq:rho:boundary} are the same in the laboratory and corotating frames, $\rho = {\tilde \rho}$. 

The absence of current density in the corotating frame, ${\tilde {\bs J}} = \sigma {\tilde {\bs E}} = 0$, implies that the rotating conductor produces a radial electric field in the laboratory frame:
\beqn
{\bs E} = - \frac{{\bs v}}{c} \times {\bs B} = - \frac{\Omega B}{c} {\bs r}_\perp.
\label{eq:induced}
\eeqn
In the last relation, we take into account the collinearity ${\bs \Omega} \| {\bs B} \| {{\bf e}_z}$, Fig.~\ref{fig:conductor}(a), and denote by ${\bs r}_\perp$ the radial component of the coordinate, ${{\bf e}_z} \perp {\bs r}_\perp$, so that ${\bs r} = {\bs r}_\perp + z {\bf e}_z$.

The rotation-induced electric field~\eq{eq:induced} generates a uniform charge density in the interior of the cylinder, $\rho = {\bs \nabla} \cdot {\bs E}/(4 \pi)$, providing us with the result~\eq{eq:rho:bulk}. The requirement of global charge neutrality leads, in turn, to the accumulation of a uniform surface charge density~\eq{eq:rho:boundary} at the edge of the cylinder. The charge density is qualitatively shown in Fig.~\ref{fig:conductor}(b). In real metals, the width $w$ of the surface layer is extremely small ($w \ll R$) being of the order of a few nanometers.

\vskip 1mm
\paragraph*{Rotation and band filling.}
Our paper is based on the simple observation that a mechanical rotation in the background of the collinear magnetic field leads to a shift of the Fermi energy $\varepsilon_F$ due to the uniform, coordinate-independent accumulation of electric charge density~\eq{eq:rho:bulk} in the bulk of the system. We analyze the situation that happens when the rotation drives the Fermi energy across the edge of the conductance or valence band. We demonstrate that this crossing breaks the discrete clockwise/counterclockwise rotational symmetry ${\bs\Omega} \to - {\bs\Omega}$ for cylinders made of a semiconducting material. The effect impacts the mechanical and conducting properties of the system.

\begin{figure}[t]
\begin{center}
\includegraphics[width=100mm, clip=true]{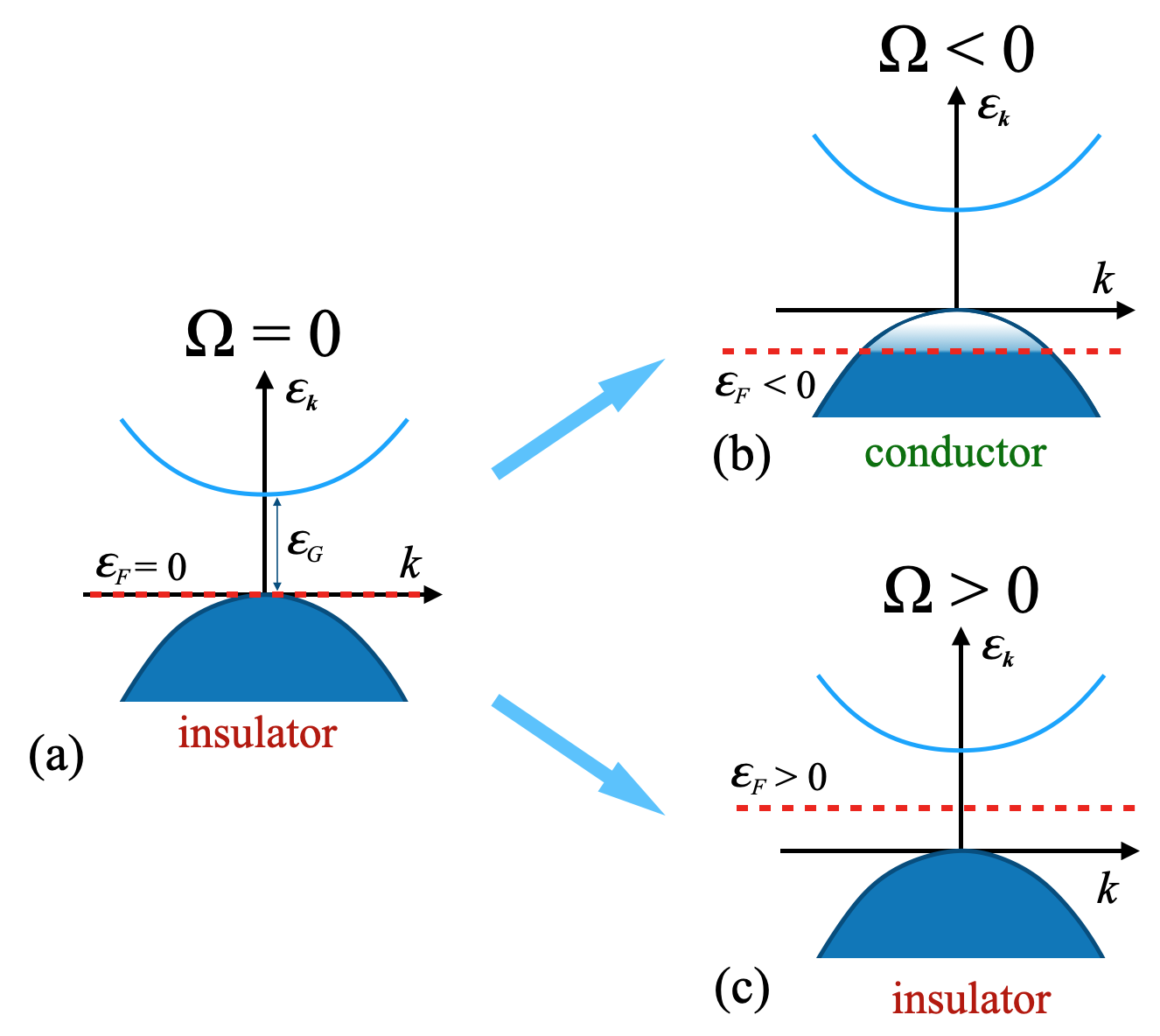}
\end{center}
%\vskip -5mm
\caption{The effect of rotation on the band filling of a semiconductor rotating with the angular velocity ${\bs \Omega} = \Omega {\bf e}_z$ in the background of magnetic field ${\bs B} = B {\bf e}_z$ with $B>0$.}
\label{fig:levels}
\end{figure}

Given the generic nature of the effect, it is sufficient to consider a degenerate semiconductor with a simple parabolic form for conduction and valence energy bands, respectively:\cite{ref:book:semiconductor}
\beqn
\varepsilon^{\mathrm{(e)}}_{\bs k} = \varepsilon_G + \frac{ {\bs k}^2}{2 m_e}, \qquad \varepsilon^{\mathrm{(h)}}_{\bs k} = - \frac{{\bs k}^2}{2 m_h}\,.
\label{eq:epsilon}
\eeqn
Here $m_e$ ($m_h$) is an effective mass of electrons (holes) and $\varepsilon_G$ is the gap between valence and conduction bands.  We neglect Zeeman and spin-orbit interactions which do not play a significant role in the effect.

\begin{figure}[t]
\begin{center}
\includegraphics[width=80mm, clip=true]{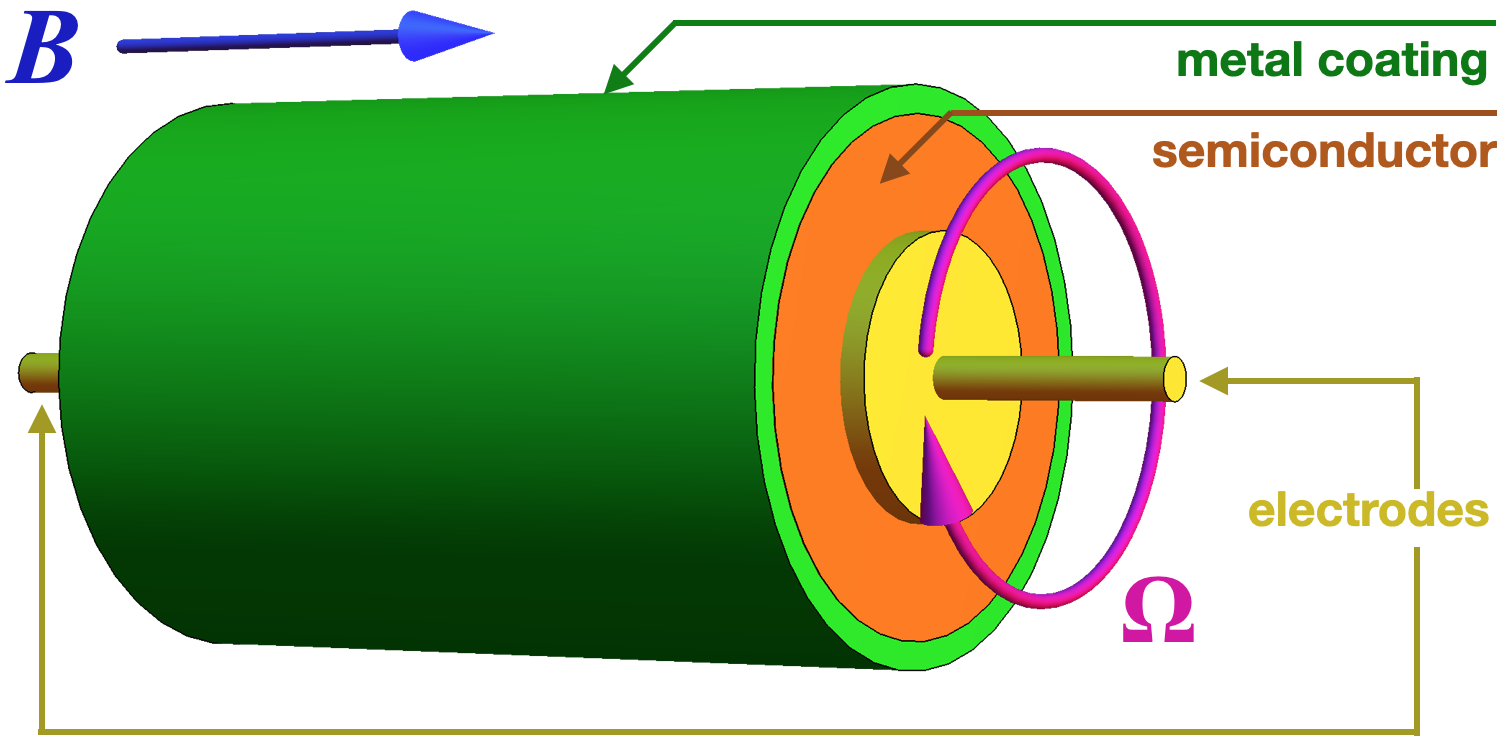}
\end{center}
\caption{The rotational diode.}
\label{fig:device}
\end{figure}

We consider a device made of an intrinsic (undoped) $p$-type semiconductor with a fully filled valence band. The Fermi energy lies in the gap close to the edge of the valence band,  as shown in Fig.~\ref{fig:levels}(a). In conventions leading to Eq.~\eq{eq:epsilon}, the Fermi energy $\varepsilon_F = 0$ corresponds to the upper edge of the valence band. We assume that the temperature is sufficiently low so that thermal energy is smaller than the energy gap between the bands, $k_B T \ll \varepsilon_G$. We also coat the cylindrical semiconductor with a thin cylindrical shell (Fig.~\ref{fig:device}) made of a metal with a wide enough conduction band that includes the Fermi energy level $\varepsilon_F = 0$. Therefore, the charge accumulation~\eq{eq:rho:bulk} happens inside the semiconducting bulk while the boundary charge buildup~\eq{eq:rho:boundary} occurs within the thin metallic layer. The electrodes are connected only to the bulk (semiconducting) part of the device and do not touch the metallic coating. 

We apply a background magnetic field along the $z$ axis, ${\bs B} = B {\bf e}_z$ (with $B > 0$) and assume that the magnetic field is sufficiently weak so that the semiconductor band spectrum~\eq{eq:epsilon} serves as a good approximation to the problem.

\vskip 1mm
\paragraph*{Conductivity and rotation.}
With the fully filled valence band, empty conduction band, and a wide energy gap between these two bands, the interior of the static (non-rotating) cylinder resides in an electrically insulating state, Fig.~\ref{fig:levels}(a). The electrons cannot be thermally excited from the valence band to the conduction band.

The clockwise rotation ($\Omega < 0$) makes the bulk charge density~\eq{eq:rho:bulk} positive implying that a part of electrons is relocated from the interior of the cylinder to its metallic boundary~\eq{eq:rho:boundary}. The external metallic coating serves as a reservoir which accommodates the electrons displaced from the interior of the system. The rotation lowers the Fermi level in the bulk of the cylinder thus creating empty states near the Fermi level. The system enters the conducting regime, Fig.~\ref{fig:levels}(b).

The shift of the Fermi energy due to rotation,
\beqn
\varepsilon_F (\Omega) = - \frac{\hbar^2}{2 m_h} \left( \frac{3 \pi \Omega B}{2 c} \right)^{2/3} \leqslant 0, \qquad  \Omega B \geqslant 0\,,
\label{eq:varepsilon:Omega}
\eeqn
is determined by a comparison of the bulk density~\eq{eq:rho:bulk} with the density of the degenerate fermionic gas (in our case, holes):
\beqn
\rho = \frac{k_F^3}{3 \pi^2 \hbar^3}, \qquad k_F = \sqrt{- 2 m_h \varepsilon_F}.
\eeqn

Thus, the clockwise rotation ($\Omega < 0$) switches the interior of the cylinder from the insulator into a conductor phase. For a slow rotation with 
\beqn
|\varepsilon_F (\Omega)| \ll k_B T \ll \varepsilon_G, 
\label{eq:hierarchy}
\eeqn
the thermally excited electrons from the Fermi sea will fill in the hole pocket in the valence band. \add{In other words, the rotation empties the energy levels in the hole pocket which can further be populated by a thermally excited electron from the Fermi sea. This process also creates a hole carrier. Thus, the electron and hole charge carriers have equal number density,} $n = p = \rho_\mathrm{bulk}/(-e)$. The conductivity of the system becomes:
\beqn
\sigma = - \frac{(\mu_e + \mu_h) \Omega B}{2 \pi c} > 0, \qquad  \Omega B < 0
\label{eq:sigma}
\eeqn
where $\mu_e$ ($\mu_h$) is the electron (hole) mobility. We used the expression~\eq{eq:rho:bulk} for the carrier concentration induced by the combined effect of magnetic field and rotation in the semiconducting bulk. Notice that the effect~\eq{eq:sigma} has a universal character in the sense that it does not depend on the details of the band structure~\eq{eq:epsilon} provided the rotation does not shift the Fermi energy~\eq{eq:varepsilon:Omega} across the boundary of the band and the hierarchy~\eq{eq:hierarchy} holds.

The counterclockwise rotation ($\Omega > 0$) induces a radial electric field which tends to make the bulk charge density~\eq{eq:rho:bulk} negative by displacing the electrons from the metallic coating to the semiconductor interior. However, the semiconducting bulk cannot accommodate them because the valence band is already fully filled, Fig.~\ref{fig:levels}(c). Therefore, the interior remains in the insulating state~\cite{footnote1}:
\beqn
\sigma = 0, \qquad  \Omega B \geqslant 0.
\label{eq:sigma:negative}
\eeqn 

%MCH zzz

The clockwise/counterclockwise asymmetry of the device is a purely quantum phenomenon based on the Pauli exclusion principle. Using symmetry arguments one shows that the effect has its roots in the absence of definite symmetry with respect to the time-reversal transformation, ${\mathrm{T}}: \, t \to -t$. Indeed, the sign flip of the (T-odd) angular velocity ${\bs \Omega}$ does not bring the device to the same state because the magnetic field ${\bs B}$ has a T-odd parity while the electric charge density $\rho$, accumulated due to the collective effect of magnetic field and rotation, is a T-even quantity:
\beqn
\mathrm{T}: \ \ {\bs B} \to - {\bs B}, \qquad  {\bs \Omega} \to - {\bs \Omega}, \qquad \rho \to \rho\,.
\eeqn

\begin{figure}[!htb]
\begin{center}
\includegraphics[width=90mm, clip=true]{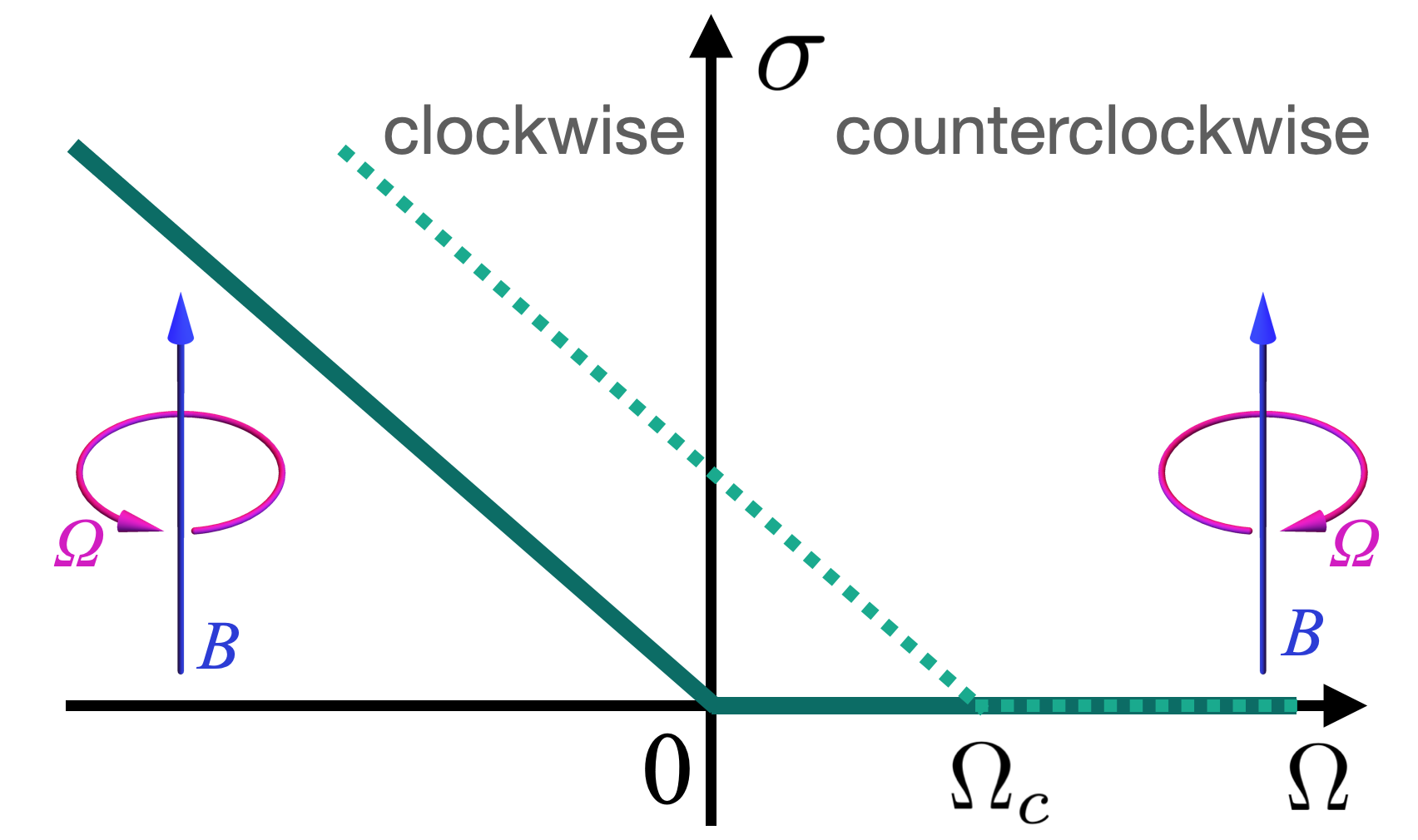}
\end{center}
\caption{The conductivity~$\sigma$ of the rotational diode vs. angular frequency $\Omega$. The solid line corresponds to a $p$-type semiconductor at the threshold Fermi level, $\varepsilon_F = 0$. The dashed line gives the generic case ($0 < \varepsilon_F <\varepsilon_G$) with a nonzero critical angular frequency $\Omega_c$, Eq.~\eq{eq:Omega:c}. The linear slopes are determined by the background magnetic field~$B >0$, Eq.~\eq{eq:sigma}.
}
\label{fig:conductivity}
\end{figure}

The dependence of the conductivity on the angular frequency, Eqs.~\eq{eq:sigma} and \eq{eq:sigma:negative}, is shown in Fig.~\ref{fig:conductivity}: the cylinder made of a semiconductor material with the threshold chemical potential behaves as an insulator for the rotation in the counterclockwise sense ($\Omega > 0$) and a conductor when it turns in the counterclockwise direction ($\Omega > 0$).  The directions are inverted with the flip of the sign of the magnetic field. 

A similar effect appears in intrinsic $n$-type semiconductors where the Fermi energy lies near the edge of the conduction band, $\varepsilon_F = \varepsilon_G$. Its phase diagram is reverted with respect to the direction of rotation, $\Omega \to - \Omega$.

Thus, the idea behind our mechanism is simple: a background magnetic field tends to shift the Fermi energy (the chemical potential) in the bulk of the rotating system thus affecting the conductivity of the latter. It can also be applied to semimetals with the Fermi energy lying in the vicinity of the upper edge of the hole pocket (the valence band) or above but close to the lower edge of the conduction band. The rotation shifts the Fermi energy across the edge of the corresponding band and thus alters its conductivity. Denoting the gap $\Delta_F = - \epsilon_F > 0$ and $\Delta_F = \epsilon_F - \epsilon_G > 0$ in the former and latter cases, respectively, we obtain that the conductor-insulator transition takes place at the nonzero critical angular frequency with the magnitude:
\beqn
\Omega_c = \frac{2 (2 m \Delta_F)^{3/2}}{3 \pi \hbar^3 B} \,,
\label{eq:Omega:c}
\eeqn
where $m$ is the mass of the appropriate carrier. The (hole) conductivity is shown in Fig.~\ref{fig:conductivity} by the dashed line.

\vskip 1mm
\paragraph*{Mechanical properties.}
The effect may also have interesting mechanical consequences. The reorganization of the electric charge density in rotating conductors produces a supplementary angular momentum $\delta {\bs L} = {\bs L}_{\mathrm{mech}} + {\bs L}_{\mathrm{e.m.}}$ which adds up to a purely mechanical quantity ${\bs L}_0$ associated with the rotation of the ionic lattice together with the original electrons. There are two contributions: the mechanical part coming from the mass redistribution of the displaced electrons ${\bs L}_{\mathrm{mech}}$ and a part originating from the angular momentum stored in the electromagnetic fields ${\bs L}_{\mathrm{e.m.}}$. 

In typical metals, the surface charge density is concentrated within a thin surface skin of a few nanometers. For practical mechanical calculations in macroscopic, centimeter-sized systems, the surface charge~\eq{eq:rho:boundary} can be treated as a $\delta$-functional distribution at the edge of the system, $r_\perp = R$. The same applies also to semiconductors, where the screening length lies in the micrometer range. Thus the electric charge density in the cylinder may be approximated by the following function:
\beqn
\rho_e({\bs r}) = \frac{\Omega B}{4 \pi c}  \left[ R \delta(r_\perp - R) - 2 \right],
\label{eq:rho:e}
\eeqn
where we used the convention $\int_0^R \delta(r_\perp - R) r_\perp d r_\perp = R$. 

The mechanical excess of angular momentum,
\beqn
{\bs L}_{\mathrm{mech}} = \int_V d^3 r \, \rho_m(\bs r) \, {\bs r} \times {\bs v}({\bs r}) = \frac{m_e L \Omega^2 R^4}{4 c e} {\bs B},
\label{eq:L:mech}
\eeqn
is determined via the surplus of the mass density
\beqn
\rho_m({\bs r}) = \frac{m_e}{e} \rho_e({\bs r})\,,
\label{eq:rho:mass}
\eeqn
where $e = + |e|$ is the elementary electric charge and $m_e$ is the mass of an electric charge carrier. Here, for simplicity, we assume the presence of a single carrier and we set its mass to the electron mass. The mechanical angular momentum~\eq{eq:L:mech} originates from the displacement of the electrons from the bulk to the boundary (or vice-versa, depending on the mutual orientation of the angular momentum~${\bs \Omega}$ and the background magnetic field~${\bs B}$). Notice that since the displaced mass is proportional to the angular frequency, Eq.~\eq{eq:rho:e} and \eq{eq:rho:mass}, the kinetic momentum~\eq{eq:L:mech} is an even function of $\Omega$.

The local angular momentum carried by the electromagnetic field $ {\bs L}_{\mathrm{e.m.}}({\bs r}) = {\bs r} \times {\bs S}$ can be expressed via the Poynting vector ${\bs S} = {\bs E} \times {\bs B}$. This calculation, however, poses a practical inconvenience as it requires the integration of the local momentum over the whole spatial volume and involves the calculation of the electric field  ${\bs E} = - {\bs \nabla} \phi$ [typically done via the electrostatic potential $\phi = \phi({\bs r})$] in the exterior of the cylinder. An equivalent definition of the electromagnetic angular momentum for spatially finite systems is based on the Maxwell form:
\beqn
{\bs L}_{\mathrm{e.m.}} = \frac{1}{c} \int d^3 r \,  \rho_e({\bs r}) \, {\bs r} \times {\bs A} = \frac{L B^2 R^4 }{8 c^2} {\bs \Omega}\,,
\label{eq:L:em}
\eeqn
which involves the vector potential ${\bs A}$ in the Coulomb gauge, ${\bs \nabla} \cdot {\bs A} = 0$. In the evaluation of Eq.~\eq{eq:L:em}, we set ${\bs B} = {\bs \nabla} \times {\bs A}$ with ${\bs A} = (B r_\perp/2) {\bf e}_\varphi$ where ${\bf e}_\varphi = {\bf e}_z  \times {\bf e}_\perp$.

The energy stored in the induced electrostatic field inside the cylinder~\eq{eq:induced}, calculated in the laboratory frame, 
\beqn
{\mathcal E}_{\mathrm{e.m.}} \equiv {\mathcal E}^{(E)}_{\mathrm{e.m.}} = \frac{1}{8 \pi} \int_V d^3 r \, {\bs E}^2({\bs r}) = \frac{\Omega^2 B^2 R^4}{16 c^2}\,,
\label{eq:E:em}
\eeqn
is related to the angular momentum~\eq{eq:L:em} via the thermodynamic relation $d {\mathcal E}_{\mathrm{e.m.}} = {\bs \Omega} \cdot d {\bs L}_{\mathrm{e.m.}}$, as expected. The energy in the corotating frame, ${\tilde {\mathcal E}}_{\mathrm{e.m.}} = {\mathcal E}_{\mathrm{e.m.}} - {\bs \Omega} \cdot {\bs L}_{\mathrm{e.m.}}$, satisfies the relation, $d {\tilde {\mathcal E}}_{\mathrm{e.m.}} = - {\bs L}_{\mathrm{e.m.}} \cdot d {\bs \Omega}$. Finally, the energy stored by the magnetic field is insensitive to rotation (up to a tiny correction due to an extra magnetic field generated via the Amp\`ere circular current).

As we discuss later, the kinetic angular momentum~\eq{eq:L:mech} associated with the displaced mass~\eq{eq:rho:mass} in non-relativistic systems is much smaller than the angular momentum stored in the electromagnetic fields~\eq{eq:L:em}. Therefore, we ignore the mechanical angular momentum in our discussion below and take for the excess of the angular momentum its electromagnetic part only, $\delta {\bs L} = {\bs L}_{\mathrm{e.m.}}$.

The electromagnetic moment of inertia ${\mathcal I}_{\mathrm{e.m.}}$ can be defined either via the angular momentum, ${\bs L}_{\mathrm{e.m.}} = {\mathcal I}_{\mathrm{e.m.}} {\bs \Omega}$, or, equivalently, via the energy: ${\mathcal I}_{\mathrm{e.m.}} = \partial^2 {\mathcal E}_{\mathrm{e.m.}}/ \partial \Omega^2$:
\beqn
{\mathcal I}_{\mathrm{e.m.}} = \frac{1}{8 \pi} \int_V d^3 r \, {\bs B}^2({\bs r}) = \frac{L R^4}{8 c^2}  B^2\,.
\label{eq:I:em}
\eeqn

In ordinary conductors, the extra angular momentum~\eq{eq:L:em} is an odd function of the angular frequency ${\bs\Omega}$: the angular momentum changes its sign, ${\bs L}_{\mathrm{e.m.}} \to - {\bs L}_{\mathrm{e.m.}}$, under the flip of the direction of rotation, ${\bs\Omega} \to - {\bs\Omega}$. The reason behind this symmetry is obvious: the rotations in opposite directions lead to the appearance of the radial electric fields of equal magnitudes (but of opposite signs) due to the displacement of electrons either from the bulk to the boundary or vice versa. The electric field emerges due to the depletion (or surplus) of the uniform electric charge density inside the bulk, depending on the direction of rotation. 

At the threshold chemical potential $\mu = 0$, the rotational diode  generates the electric field for the particular rotational direction for which ${\bs \Omega} \cdot {\bs B} <0$. The rotation in opposite direction (${\bs \Omega} \cdot {\bs B} >0$) does not produce the electric field in the bulk. Therefore, the rotational energy and angular momentum stored in the induced electric field differs for the rotations in the clockwise (CW) and counterclockwise (CCW) senses. For the device at the threshold value of the Fermi level, the difference between the angular momenta for the clockwise/counterclockwise rotations is given by Eq.~\eq{eq:I:em}:
\beqn
\Delta {\mathcal I}_{\mathrm{e.m.}} \equiv {\mathcal I}^{\mathrm{CW}}_{\mathrm{e.m.}} - {\mathcal I}^{\mathrm{CCW}}_{\mathrm{e.m.}} = - \frac{L R^4}{8 c^2}  B^2 \, \mathrm{sign}\left( {\bs \Omega} \cdot {\bs B} \right). \quad
\label{eq:Delta:I}
\eeqn
Interestingly, this quantity has a universal character in a sense that it depends only on the geometry of the device and the background magnetic field.

While the odd nature of the effects and the simplicity of the device that hosts them may seem attractive, quantitative estimates, given below, challenge the suitability of these effects for an experimental detection since their magnitude is not exceptionally large. 

\vskip 1mm
\paragraph*{Electric charge in the bulk.}
In the background of the moderate magnetic field~$B = 1\,T$, the cylinder rotating with the angular frequency\cite{footnote4} $\Omega = 100 \, \mathrm{s}^{-1}$ accumulates in its interior the electric charge with the bulk density $\rho_{\mathrm{bulk}} = - 8.8\times 10^3\, e/\mathrm{cm}^3$. This is a small, but still non-negligible number. 

\vskip 1mm
\paragraph*{Conductivity.}
The rotation-induced conductivity~\eq{eq:sigma} depends substantially on the mobility $\mu$ of the charge carriers. For a typical semiconductor with mobility $\mu_e \sim \mu_e \sim 10^3$\,cm${}^2$/(V$\cdot$s), an $\Omega = 100 \, \mathrm{s}^{-1}$ rotation in the magnetic field background $B = 1\,T$ induces conductivity $\sigma \simeq 10^{-13} \, \mathsf{\Omega}^{-1} \mathrm{cm}^{-1}$ which makes the interior of the device as ``good'' conductor as, for example, glass or rubber. This exceptionally bad conductivity can, however, be improved in systems with a high mobility of charge carriers (possibly gated to achieve the correct position of the Fermi level at the upper/lower edge of a hole/electron band so that the non-rotating system resides at the border of an insulating state). For example, for AlGaAs/GaAs heterostructures featuring the high-mobility two-dimensional electron gas with\cite{ref:AlGaAs} $\mu = 3.5\times 10^7$\,cm${}^2$/(V$\cdot$s), the rotation induces conductivity compared to the lowest value $\sigma \simeq 10^{-8} \, \mathsf{\Omega}^{-1} \mathrm{cm}^{-1}$ achievable in a pure semi-insulating GaAs crystal\cite{ref:GaAs}.

\vskip 1mm
\paragraph*{Angular momentum.}
The kinetic angular momentum associated with the radial displacement of electrons~\eq{eq:L:mech} is much smaller than the angular momentum stored in the electromagnetic field~\eq{eq:L:em}: $L_{\mathrm{mech}}/L_{\mathrm{e.m.}}  \sim 10^{-9}$. Therefore, the mechanical properties of the device are determined only by the electromagnetic fields generated by the rotation. The difference between the angular moments of inertia for clockwise and counterclockwise rotation~\eq{eq:Delta:I} of a centimeter-sized cylinder ($L = R = 1 \,\mathrm{cm}$) in the magnetic field $B = 1\,T$ is $\Delta {\mathcal I}_{\mathrm{e.m.}} \simeq 10^{-15} \, \mathrm{g}\cdot \mathrm{cm}^2$. No matter how small this number may seem from the first sight, it corresponds to a moment of inertia of a water droplet in a typical fine fog (with the size about $10 \mu$m) which is already a macroscopic object. The latter example gives us some hope that the change in the clockwise/counterclockwise moments of inertia may be within experimental reach despite it constitutes a negligible fraction of the total moment of inertia of the system, $\Delta {\mathcal I}_{\mathrm{e.m.}} / {\mathcal I}^{\mathrm{tot}}_{\mathrm{e.m.}} \simeq 10^{-16}$.

\vskip 2mm
\paragraph*{Summary.}
We demonstrated that the absence of a definite time-reversal state in a mechanically rotating semiconductor in the background magnetic field leads to asymmetry of its mechanical and transport properties with respect to rotations in clockwise and counterclockwise directions. A fine-tuned system becomes a ``rotational diode'' that possesses different moments of inertia and resides in different conductor/insulator phases when rotated in opposite directions. Although the effect bears its roots in classical electrodynamics of rotating conductors in the background magnetic field, the clockwise/counterclockwise rotational asymmetry appears as a purely quantum phenomenon based on the Pauli exclusion principle associated with the Fermi-Dirac statistic of electrons. The estimated magnitude of these effects turns out to be rather small for realistic materials.

%\acknowledgments
\clearpage\paragraph*{Acknowledgments.}
The author is grateful to Karl Landsteiner and Mar\'ia Vozmediano for correspondence and interesting discussions.
The work is supported by Grant No. 0657-2020-0015 of the Ministry of Science and Higher Education of Russia.


\begin{thebibliography}{99}
\bibitem{footnote1}
An exceptionally fast rotation with the energy scale of the interband gap energy can produce a sufficiently strong radial electrochemical potential, $\phi = - \Omega B r_\perp/c$ which can gradually fill in the conduction band provided $\varepsilon_G - \phi(L) {>} 0$.
\bibitem{footnote4} It is of the order of the angular frequency of the consumer hard disk,  $\Omega = 7200 \,\mathrm{rpm} \sim 10^{2}\,\mathrm{s}^{-1}$.
\end{thebibliography}
\end{document}